\documentclass[aps,preprint]{revtex4}
\usepackage[dvips]{graphicx}

\begin{document}

\title
{
Vortex motion  
in chilarity-controlled pair of magnetic disks
}

\author
{H. Masaki, T. Ishida}
\affiliation{
Institute for Solid State Physics, University of Tokyo, 
5-1-5 Kashiwanoha, Kashiwa, Chiba 277-8581, Japan 
}

\author
{R. Antos, J. Shibata}
\affiliation{
RIKEN FRS, 2-1 Hirosawa, Wako, Saitama 351-0198, Japan 
}

\author
{ 
T. Kimura, Y. Otani 
}
\affiliation{
Institute for Solid State Physics, University of Tokyo, 
5-1-5 Kashiwanoha, Kashiwa, Chiba 277-8581, Japan \\
RIKEN FRS, 2-1 Hirosawa, Wako, Saitama 351-0198, Japan 
}

\date{\today}
\begin{abstract}
We investigate the influence of the vortex chirality on the magnetization processes of a magnetostatically coupled pair of magnetic disks.  
The magnetic vortices with opposite chiralities are realized by introducing asymmetry into the disks.  
The motion of the paired vortices are studied by measuring the magnetoresistance 
with lock-in resistance bridge technique.  
The vortex annihilation process is found to depend on the moving directions of 
the magnetic vorticies.  The experimental results are well reproduced by the micromagnetic simulation.

\end{abstract}

\maketitle

Patterned domain structures are useful for the application
in the future spintronic devices as well as further understanding
of the fundamental spin-related physics.  
Especially the magnetic vortex structure,
stabilized in a ferromagnetic disk
with a diameter less than a micron,
has a potentiality as a unit cell of
high density magnetic storage because of negligible magnetostatic interaction.\cite{Cowburn, Shinjo}  
Therefore, understanding the magnetization process of the magnetic vortex
is an important issue for further development.  
In a single magnetic disk, the magnetic properties
such as the susceptibility and the field stability can be determined
by geometrical parameters, the aspect ratio of the diameter to the thickness.\cite{Gus1, Gus2}   
In two or more magnetic disks,
 another geometrical parameter, the separation distance between disks,
becomes important because of the magnetostatic interaction.\cite{Nov}  
Although the magnetic vortex at the remanent state does not induce the
magnetostatic interaction because of the flux-closure domain,
application of the in-plane magnetic field breaks the symmetry in the flux-closure structure,
leading to the magnetic charge at the edge of the disk.  
When the separation distance is much smaller than the diameter,
the magnetic properties are strongly modified.   
This means that the magnetic properties of the vortex can be tuned
by adjusting the separation distance.

The magnetic vortex is described by two topological quantities; polarity and chirality.  
The vortex polarity, the magnetization direction of the vortex core,
strongly correlates to the dynamical trajectory of the vortex core\cite{Gus2, Shibata1}
and the spin-current induced vortex motion.\cite{Shibata2, Ishida}
The chirality, the rotational direction of magnetic moments
whirling either clockwise (CW) or counterclockwise (CCW),
determines the direction of the vortex shift induced by the in-plane magnetic field.  
In a single magnetic disk, these two quantities do not contribute to
the magnetization process because of the well defined symmetry.  
However, in coupled vortex systems,
these quantities are expected to affect the vortex behaviors because of the magneto-statically induced asymmetric interaction.  
Here, we study the influence of the vortex chirality in a system of two coupled vortices.  
We found that the vortex chirality strongly affects the vortex annihilation fields.

In a perfect circular disk, CW and CCW states are degenerate 
because the in-plane shape anistropic energy is isotropic.  
However, the chirality of the magnetic vortex is known to be controlled by introducing the asymmetry.  
For example, in the disk with a flat edge on one side,  
the vortex easily nucleates from the flat edge assisted by the larger
demagnetizing field than that at the round edge
when the magnetic field is applied parallel to the flat edge.\cite{Sch}  
Thus, the chirality can be tuned by the magnetic field.  
We consider paired magnetic disks in which each disk has a flat edge
at left- or right-hand side, as shown in Fig.\ 1(a).  
When the positive vertical magnetic field is applied,
the vortex states with opposite chiralities (disk A : CW, disk B : CCW)
are stabilized in the remanent state.  
In such a vortex system,
the horizontal magnetic field induces
a mutually opposite vortex movement.  
When the positive horizontal field is applied in the remanent state,
the vortices approach each other with increasing magnetic field(i.e. inward motion).  
On the other hand, when the negative horizontal magnetic field is applied,
the vortices move away from each other(i.e. outward motion).  
These two situations may cause the difference in the magnetic interaction.

The magnetically coupled pair of asymmetric disks
are prepared by means of conventional electron-beam lithography and lift-off technique.  
Py disks 40 nm in thickness were grown using an electron beam evaporator
with a base pressure of 2 $\times 10^{-9}$ Torr.   
Figure 2(a) shows a scanning-electron-microscope (SEM) image of a prepared device.  
The diameter of the disks are 1 $\mu$m and the separating distance is 200 nm.  
5 percents of its diameter at the right and left sides are
cut in disk A and disk B, respectively, to induce the mutually opposite chirality.   
When the initial vertical positive magnetic field more than 1000 Oe is applied,
the CW and CCW chiralities are respectively formed in disks A and B.

The vortex motion can be detected by measuring the magnetoresistance (MR).  
Two Cu probes are connected to each disk for the MR measurements.   
The resistance bridge method combined with the lock-in detection is employed as 
schematically shown in Fig.\ 2(a), 
to remove the background resistances such as contact, electrode and Py resistances
in two terminal measurements.  
In the vertical magnetic field,  
the vortex nucleates from the flat edge, then moves to the round edge and then annihilates.  
As in Fig.\ 2(b), the change due to anisotropic MR
is clearly observed with the two jumps corresponding to
the vortex nucleation and annihilation respectively at $-100$ Oe  and $360$ Oe.

In order to confirm that the magnetic vortex has a desired chirality, 
the positional dependence of the vortex annihilation field is measured.  
When the vertical magnetic field is swept from the negative to the positive directions 
as in the inset of Fig.\ 2(b),
the magnetic vortex nucleates from the flat edge and annihilates from the round edge.  
However, when the field sweep direction is reversed after the vortex nucleation,
the magnetic vortex goes back to the flat edge and then annihilates from the flat edge.  
Figure 2(c) shows the MR when the magnetic field is swept from 0 Oe to $-1000$ Oe 
after the vortex nucleation.  
The vortex annihilation is observed at $-420$ Oe 
which is larger than that from the previous measurement.  
The similar tendency is observed in disk B.  
According to the micromagnetic simulation using the object-oriented
micromagnetic framework (OOMMF) software,\cite{OOMMF}
the annihilation field from the flat edge is 500 Oe larger
than 450 Oe corresponding to the annihilation field for the round edge.  
Here, the cell size, the saturation magnetization $M_S$
and the damping parameter $\alpha$ are 10 nm, 1.0 T
and $\alpha = 0.5$, respectively.  
These facts clearly indicate that the vortex chirality has a desired direction.

As mentioned above,
the paired magnetic vortices have mutually different chiralities 
in the remanent state after applying the positive vertical magnetic field of 2000 Oe.   
We here study the difference between the inward and outward motions of paired vortices.  
Figure 3(a) shows the resistance change of the disk A as a function of 
the horizontal positive magnetic fields inducing inward vortex motions.  
The vortex annihilation is clearly observed as a step in the resistance change at 500 Oe.  
Figure 3(b) shows the resistance change as a function of 
the horizontal negative magnetic fields inducing outward vortex motions.  
The annihilation field is 550 Oe, larger than that in inward vortex motion.  
Thus, the difference of the annihilation field between two measurements are clearly observed.  
The vortex annihilation fields both for the inward and outward motions 
are larger than that for the vertical field.  
This is because the vertical axis corresponds to a easy axis for the paired vortex system.\cite{Nov}

To explain the difference in the annihilation field between the two vortex motions,
we simply attribute to the different magnetostatic interaction for above two cases.  
When the vortices move outward, the magnetic charges appear at inner side
of the coupled disks and increasing horizontal magnetic field as in Fig.\ 4(a).   
Such magnetic charges develop strong magnetostatic interaction and
prevent the vortex motion to outside.  
On the other hand, when the vortices move inward,
the magnetic charge mainly appears at the outer side of the coupled disks.  
As in Fig.\ 4(b), when compared, the magnetostatic interaction is not so strong in the inward motion.  
In this way, the annihilation field for the outward vortex motion
is larger than that of the former situation.

To understand more quantitatively, we also studied the vortices' motions with the mutually 
opposite chiralities under similar magnetic field conditions by micromagnetic simulations 
using the same parameters as the previous calculations.  
We found that the annihilation field for the inward motion is 600 Oe, 
smaller than $-700$ Oe for the outward motion.  
Thus, the experimental results are well reproduced by the micromagnetic simulations.

In conclusion, we study the influence of the vortex chirality in the coupled pair of vortices.  
The annihilation field is found to depend on the vortex chirality
when the two vortices have mutually opposite chiralities.  
This result indicates that 
the vortex chirality is an additional parameter for 
controlling the magnetic property in the coupled vortex systems.


\newpage

\begin{figure}
\caption
{
(a) Formation of the magnetic vorticies with mutually opposite chirality 
by introducing the positive vertical magnetic field $H_{\rm ver}$.  
After the application of the magnetic field, 
the chiralities of disk A and disk B are CW and CCW, respectively.  
(b) Inward vortex motion by applying positive horizontal magnetic field $H_{\rm hor}$.  
(c) Outward vortex motion by applying negative horizontal magnetic field.  
}

\caption
{
(a) Schematic diagram of the bridge circuit for the 
magnetoresistance measurement together with a SEM image of the 
fabricated paired asymmetric disks.  
(b) Resistance of disk A as a function of the vertical magnetic field 
together with the corresponding domain structures calculated by micromagnetic simulations.   
(c) Resistance of disk A as a function of the vertical magnetic field 
together with the corresponding magnetic structures.  
After the nucleation of the vortex, the magnetic field is swept from 0 
back to $-1000$ Oe.  
}

\caption
{
(a) Resistance of disk A as a function of the positive horizontal magnetic field.  
The vortices approaches with increasing the magnetic field.  
(b) Resistance of disk A as a function of the negative horizontal magnetic field.  
The vortices go away with increasing the negative magnetic field.  
The arrows indicate the vortex annihilations.  
}

\caption
{
Schematic illustration of the magnetostatic interaction 
(a) at inward motion and (b) at outward motion.  
}

\end{figure}

\vspace*{1cm}
\newpage

\newpage
\vspace*{2cm}
\begin{center}
\includegraphics[scale=0.6]{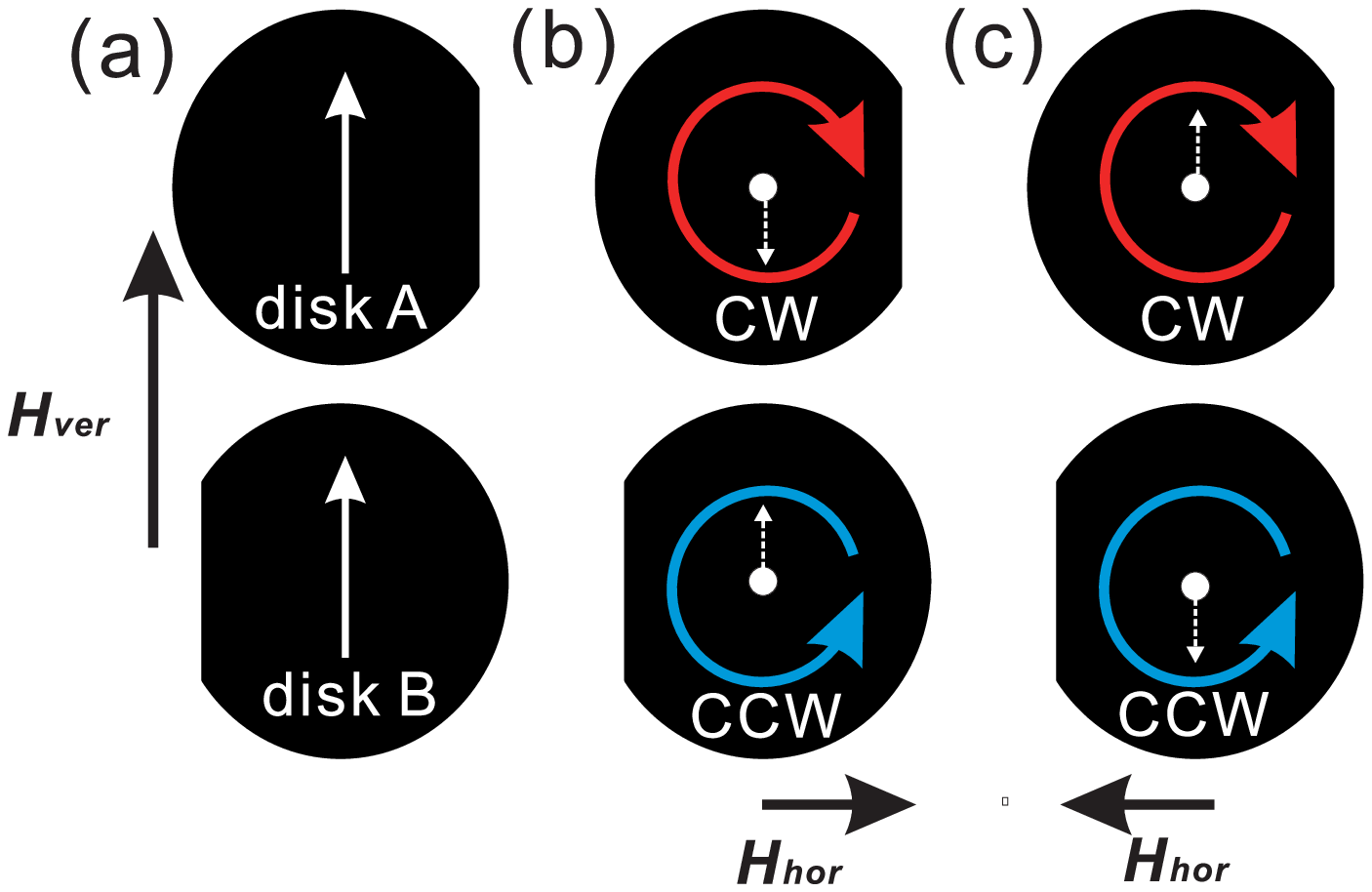}
\end{center}
\vspace*{1cm}
\begin{center}
Fig.\ 1 Masaki et al. \\ 
Color only in on-line
\end{center}

\newpage
\begin{center}
\vspace*{3cm}
\includegraphics[scale=0.6]{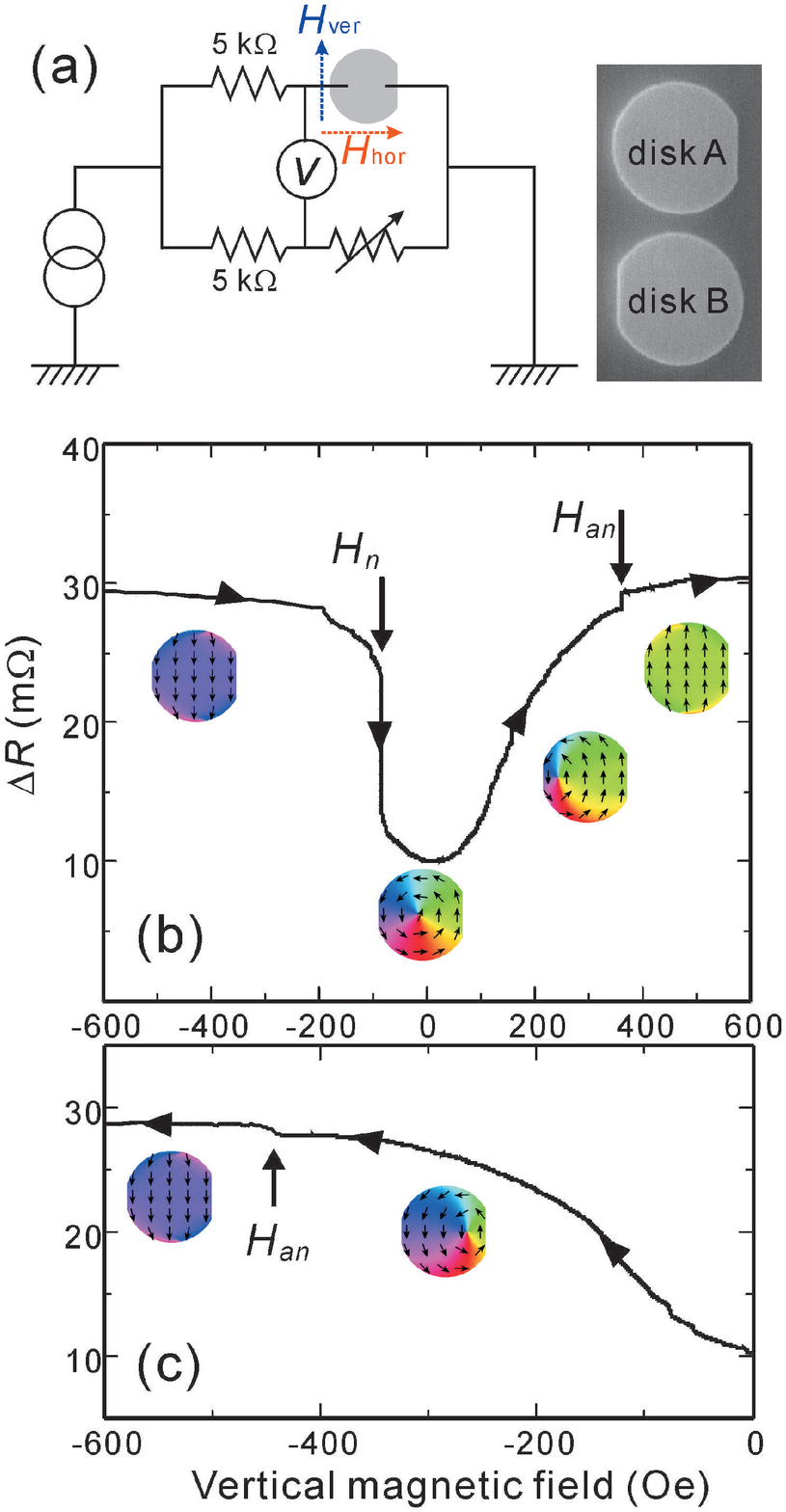}
\end{center}
\vspace*{1cm}
\begin{center}
Fig.\ 2 Masaki et al. \\
Color only in on-line
\end{center}

\newpage
\begin{center}
\vspace*{3cm}
\includegraphics[scale=0.6]{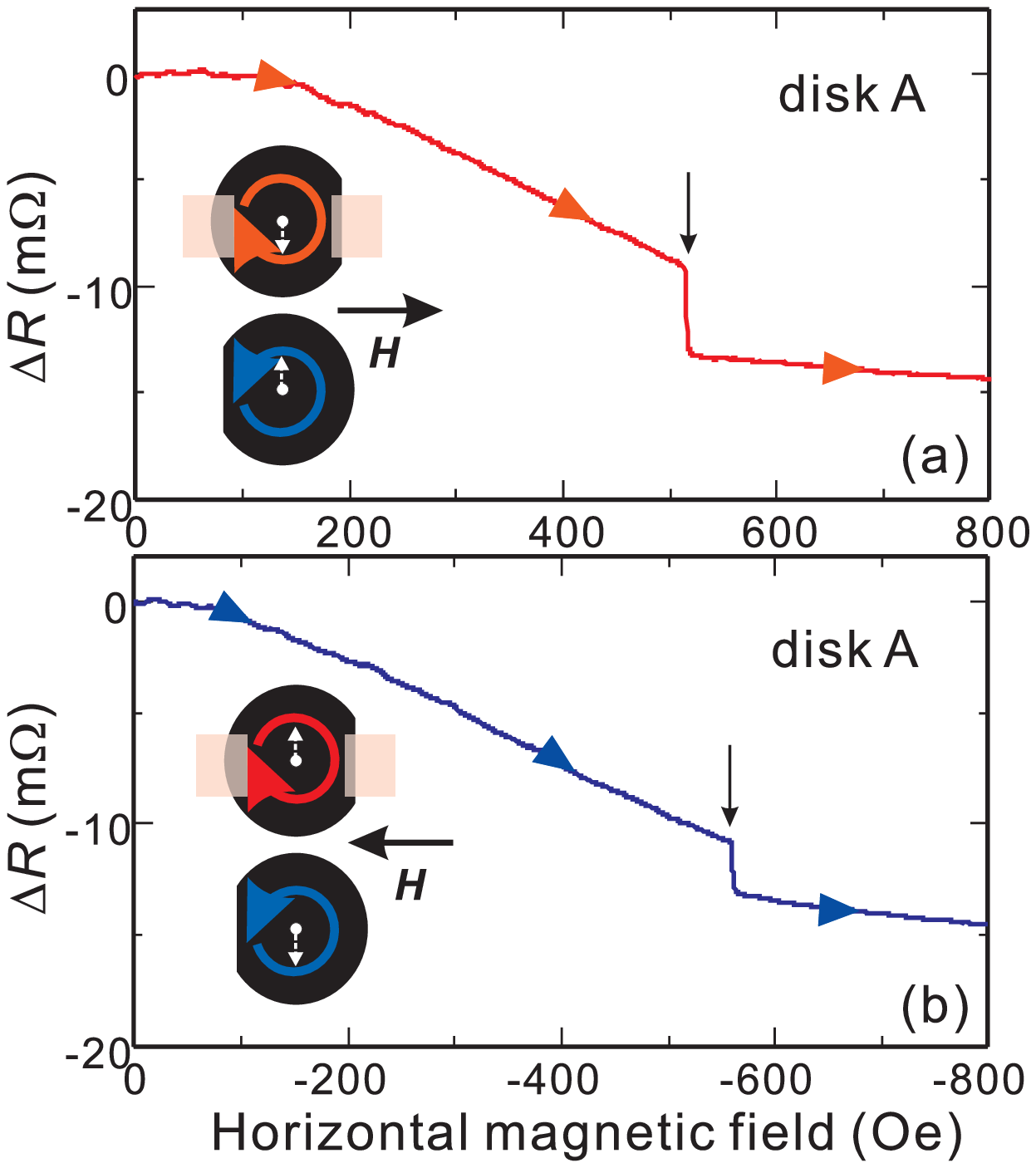}
\end{center}
\vspace*{1cm}
\begin{center}
Fig.\ 3 Masaki et al. \\
Color only in on-line
\end{center}

\newpage
\begin{center}
\vspace*{3cm}
\includegraphics[scale=0.8]{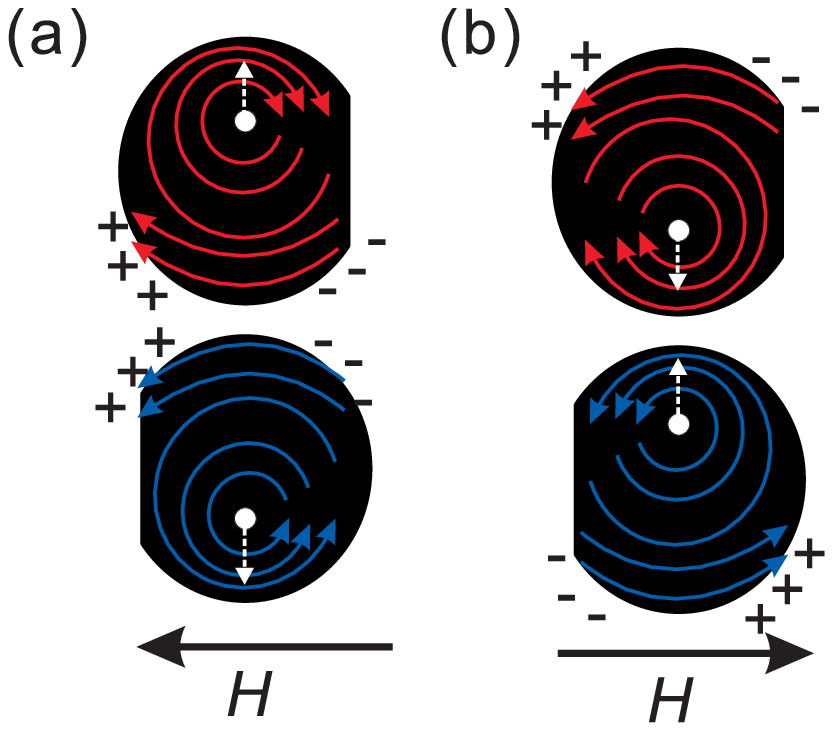}
\end{center}
\vspace*{1cm}
\begin{center}
Fig.\ 4 Masaki et al. \\
Color only in on-line
\end{center}

\end{document}